\documentstyle[11pt, twoside, paspconf, epsf]{article}

\begin{document}

\input{psfig.tex}

\title{BL Lac Objects and Blazars: Past, Present, and Future}

\author{C. Megan Urry\altaffilmark{1}}
\affil{Space Telescope Science Institute, 3700 San Martin Drive,
	Baltimore MD 21218}
\altaffiltext{1}{Sabbatical visitor at the Center for Astrophysical
Sciences, The Johns Hopkins University}



\begin{abstract}
The past 20 years have seen phenomenal progress in our understanding
of BL Lac objects. They form part of the blazar class, which are
radio-loud AGN whose relativistic jets are aligned along our
line of sight. Several critical milestones have helped establish
this picture, first proposed at the Pittsburgh BL Lac meeting 20 years
ago, most recently the EGRET and TeV detections of beamed gamma-ray
emission. The spectral energy distributions are double peaked and
follow a self-similar sequence in luminosity, which can be explained
by electron cooling on ambient photons. This simple paradigm has
yet to be tested, and further questions remain, notably about
physical conditions in blazar jets --- the kinetic power, magnetic
energy density, acceleration time scales, proton content, etc. ---
and how this energy is transported in the innermost regions. Some clues
are available from multiwavelength monitoring campaigns although better 
sampling over longer periods is clearly called for. 

Recent work on the host galaxies of BL Lac objects supports 
their unification with low-power, aligned FRI radio galaxies. 
Nature makes jets with a large range of kinetic powers, but the
distribution of such jets --- i.e., the relative
number densities of low-luminosity (``blue'' BL Lacs) or high-luminosity
(``red'' BL Lacs and FSRQ) blazars --- is highly uncertain. Deeper
radio-selected samples of flat-spectrum sources may resolve the degeneracy
in the present demographic analyses.

Since according to unified schemes blazars are representative
of all radio-loud AGN, their jet properties have broad implications.
Future EUV/X-ray observations will illuminate the circumnuclear structure,
especially the hot, highly ionized, high velocity gas
on sub-parsec scales, which could play a role in jet dynamics
and could possibly affect the formation of FRI vs. FRII type jets.
The study of blazars may also help us eventually understand the
difference between radio-quiet and radio-loud AGN.

\end{abstract}



\section{Legacy of the BL Lac Conferences}

The Turku meeting was the third in an historic series of 
conferences on BL Lac objects, so it seems appropriate to begin
from an historical perspective. The first BL Lac meeting was held in 
Pittsburgh in 1978 and the second 10 years later, in Como, Italy.
Over this 20-year period, there has been enormous progress in 
understanding BL Lac objects, or blazars, using the collective term 
for BL Lac objects and flat-radio-spectrum, highly polarized quasars.
This occurred despite the scarcity of sources --- in 20 years
the number of catalogued BL Lacs increased only by a factor of a few,
roughly 1/10 the growth in catalogued quasars. (We have at least reached the 
point where the number of BL Lac objects is greater than the number
of conference participants!)

The field has changed a great deal in 20 years. 
While many of the 84 Pittsburgh participants have gone on to illustrious 
careers in astrophysics (this is the good news, for students who wonder 
whether blazars are a good career move), almost none are still actively 
working on blazars (the bad news?). For example, only 8 people from the 
Pittsburgh meeting were at the 1988 Como meeting, and only 1 person 
came to both Como and 
Turku (answer at the end). This represents real turnover, as the number
of blazar researchers has not increased dramatically --- roughly the
same number of people came to Como as Pittsburgh. In Turku the attendance
was 50\% higher, in large part because of the welcome
influx of gamma-ray astronomers.
Altogether, the demographics of these three conferences indicates 
considerable evolution in our understanding and renewed interest in
the physics of blazars. 

In his overview at Pittsburgh, Wayne Stein was reliably prescient. He
suggested tongue-in-cheek that BL Lac objects might be defined as
``totally uninteresting astronomical sources [to the spectroscopist]
because they may exhibit neither absorption nor emission lines.'' (This
surely resonates with many optical observers who have trouble getting telescope
time for spectroscopy!) He also pointed out, more seriously, the reason
for the high level of interest in this topic, saying,
``BL Lac objects are our most direct observable link with the ultimate
energy source of the quasi-stellar objects.'' Indeed, we now recognize that
BL Lac objects and other blazars are the best probe of jet formation and 
evolution in AGN because, although they are rare, unification tells us 
they are the same as all radio-loud AGN, and at the same time, their 
jet emission is conveniently enhanced by relativistic beaming. 
Thus blazars are critical to understanding jets in all radio-loud AGN --- 
what was discussed at Turku about BL Lac objects has near-universal 
relevance for AGN.

We can look back at what have we learned since Pittsburgh, what we knew then,
what have been the major problems and solutions, and what are the most 
promising avenues for progress in the future.
Over 10-year intervals, not surprisingly, real progress does happen! 
But it's also humbling to recognize how much was already known in 1978.
In the most quoted paper of any conference proceedings ever, 
Blandford and Rees (1978) already outlined a fair fraction of 
the underlying physics of BL Lac objects:
they understood that the spectral energy distributions (SEDs) were dominated
by synchrotron and probably inverse Compton emission;
they realized the emission region(s) had to be relativistically expanding; 
and they even knew that the selection effects would favor the discovery of 
BL Lac objects even though they were rare objects. 

It is also interesting to observe the evolution of topics under discussion
at Pittsburgh, Como, and Turku:
\begin{itemize}
\item {\it Observational properties at various wavelengths ---}
There was the usual break\-down
into wave\-lengths --- ra\-dio, op\-ti\-cal, X-ray --- but the mix changes.
In Turku there was more 
discussion of VLBI results, and certainly more about gamma-rays. 
The discovery of strong gamma-ray emission from blazars has completely
changed our understanding of their energy output.
\item {\it Multiwavelength monitoring ---}
The topic of multiwavelength monitoring first came up in Como and 
was discussed much more in Turku. Spectral variations are a very powerful 
tool for understanding the physics of BL Lac objects. 
\item {\it Spectroscopy: line strengths and redshifts ---}
In Pittsburgh, there was a great deal of interest in spectral lines and
distances. These are still important, despite being less discussed today.
In one of our most important BL Lac samples, the 1~Jy radio-selected
sample, about 1/4 of the redshifts are still missing.
\item {\it Models ---}
Theorists continue to keep up, and often to be ahead of the observations.
We now understand a lot about the physics of blazars and can even 
speculate on the similar physics in Galactic black hole candidates and 
perhaps gamma-ray burst sources.  
\item {\it Surveys and population statistics ---}
The small number of well-defined blazar samples indicates the critical
importance
of surveys. In Como, the first flux-limited samples were presented; these
were essential for understanding the true nature of BL Lac objects. 
In Pittsburgh, there was one talk on number counts (Setti 1978); 
in Como, there were more, plus speculation about luminosity functions, 
evolution, and unification.
Since then, unification has been put on a firm statistical basis using 
the new flux-limited samples (see Padovani, this volume).
\item {\it Gamma-rays ---}
Another significant advance since Como was the discovery of 
blazars as bright gamma-ray sources, first at GeV energies with CGRO
EGRET and more recently at TeV energies with Whipple and subsequent
ground-based Cerenkov detectors. A large fraction of blazar light 
had previously been missed. The highly variable, strong gamma-ray 
emission confirms that blazars are relativistically beamed and 
offers strong constraints on the physics of the jet.
\item {\it Host galaxies ---}
There is considerable new data on host galaxies of BL Lac objects,
from HST and ground-based observatories.
In Pittsburgh there was speculation about whether BL Lacs sit in galaxies;
it is now clear they are the active nuclei of normal, fairly luminous
elliptical galaxies with intrinsically low radio powers.
\end{itemize} 

In this talk I review in broad terms our current understanding of 
BL Lac objects, and point out where progress is being made or where it 
can be made. Most of these topics are discussed in more detail throughout
this volume.

\section{Spectral Energy Distributions}

The spectral energy distributions (SEDs) of blazars have a universal
two-bump structure (in $\nu F_\nu$, Fig. 1), almost certainly caused by
synchrotron radiation peaking at low energies, and at gamma-ray energies,
possibly inverse Compton-scattering of soft
photons radiation (although there are other possibilities; see below). 
While I may refer to the gamma-ray component as the ``Compton component'' 
and discuss the implications of inverse Compton scattering models, 
bear in mind that this model is not yet established. 

The characteristics of blazar SEDs, and particularly their spectral
variability, offer clues to the underlying physics in blazar jets.
(Here ``jet'' means the unresolved, inner jets of radio-loud AGN,
which dominate the observed SED. The longer-wavelength radio
emission comes from more extended regions of the jet, 
which are sometimes resolved in radio or even optical images.)

\subsection {``Red'' and ``blue'' blazars}

For more than a decade, BL Lac objects have been sorted into two types
according to their broad-band spectral properties (e.g., 
$\alpha_{ro}$ vs. $\alpha_{rx}$; Ledden \& O'Dell 1985, Stocke et al. 1985);
these can be described as
``red'' BL Lacs, technically defined as LBL (Low-frequency
peaked BL Lacs) and found primarily in radio-selected samples,
and ``blue'' BL Lacs, a.k.a. HBL (high-frequency
peaked BL Lacs), found primarily in X-ray-selected samples. 
(Strong emission-line blazars [quasars] are very similar to red BL Lacs.)
It is now becoming clear that the SEDs, luminosities, emission line and 
other properties of blazars form a continuous distribution,
and that the original red-blue dichotomy was caused by selection biases 
at high flux limits
(see talks by Laurent-Muehleisen, Padovani, Ghisellini, and others, 
this volume).

\begin{figure}
\centerline{\psfig{file=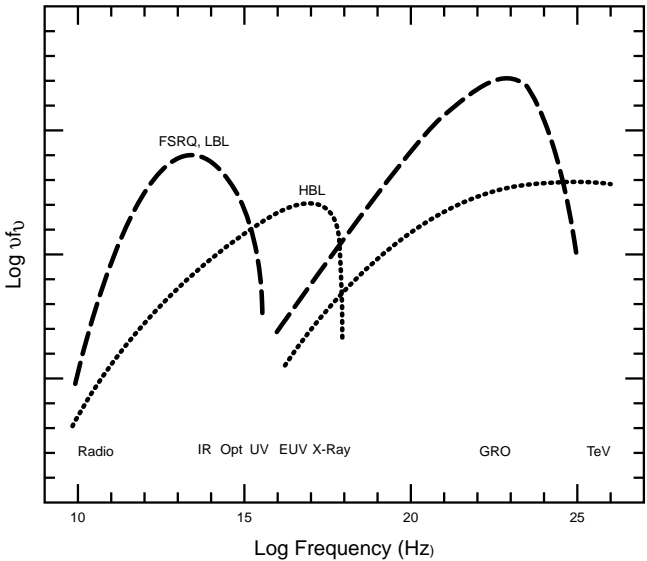,width=0.6\linewidth}}
\caption{Characteristic double-peaked SEDs of blazars. The 
low-energy component, due to synchrotron radiation, peaks in the
infrared-optical for ``red'' blazars (dashed line; FSRQ,LBL) 
and at UV-X-ray energies for ``blue'' blazars (dotted line; HBL). 
The corresponding high-energy components, which may be due to 
inverse Compton scattering of soft photons, peak at GeV or TeV energies, 
respectively.
\label{fig:schematic}
}
\end{figure}
\normalsize

\begin{figure}
\centerline{\psfig{file=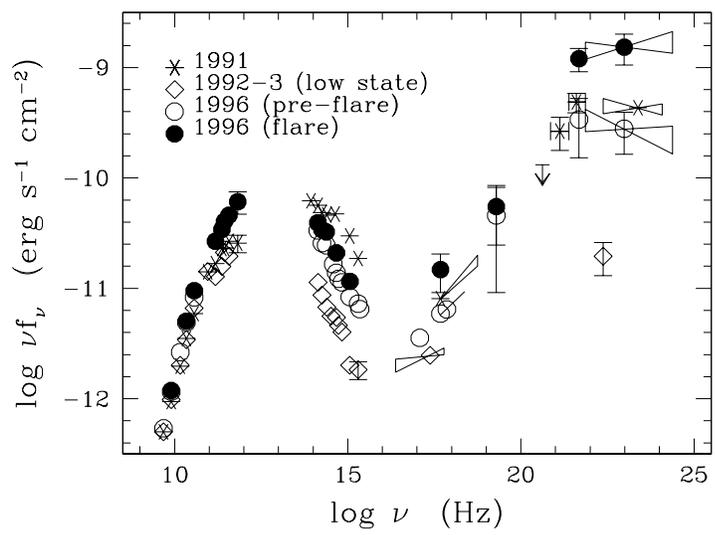,width=0.7\linewidth}}
\caption{Spectral energy distribution of 3C279 at multiple epochs
spanning flare to faint states. Much larger amplitude variability is seen 
in the gamma-ray component than in the synchrotron component, and the
amplitude is larger near the peak. 
\label{fig:3c279}
}
\end{figure}
\normalsize

\subsection {Variability}

Blazars vary most at frequencies above the peak in both SED components,
and intensity variations are approximately correlated between 
the two peaks (Fig.~2; also, Ulrich, Maraschi \& Urry 1997).
The lags at high energies are short --- typically less than a day --- 
and depend on wavelength. The variability amplitudes are usually 
larger at higher energies, increasing greatly above the SED peak
frequencies. 
The correlated variability suggests that the
same electrons could be radiating both components.

It has long been noted that blue BL Lacs are much less variable 
in the optical than red BL Lacs. This has been taken to mean that
they are somehow less active. Instead, from
an SED-based perspective, the difference is an accident of where
the SED peak occurs relative to the optical band:
in red BL Lacs, optical frequencies lie above the synchrotron peak,
the region of maximum variability, whereas in blue BL Lacs this
is well below the synchrotron peak frequency, where variability 
is typically small. 
In fact, the blue BL Lacs are at least as variable, 
in terms of large amplitude and short time scale, as red BL Lacs,
but at the corresponding part of the SED, i.e., at X-ray energies. 
By the same argument,
one would predict high EUV/X-ray polarization in blue BL Lacs,
comparable to the optical polarization in red BL Lacs.

\subsection{Synchrotron and Inverse Compton-scattering emission models}

Many speakers address inverse Compton scattering
(Ghisellini, Kirk, Marscher, Levinson, this volume). The various models
share several basic characteristics, including
a region of energetic particles moving relativistically, 
often in an expanding jet geometry commensurate
with the unification scheme for blazars (Urry \& Padovani 1995).
These particles radiate synchrotron radiation and also
inverse Compton scatter ambient photons to gamma-ray energies. The ambient
photons can be synchrotron photons in the jet
(the synchrotron self-Compton, or SSC, model), 
or UV/X-ray photons from the disk, 
or reprocessed UV (or even IR) photons from the broad-line region
or obscuring region beyond. 
Which of these soft photon sources dominates depends on their local
energy density near the electrons of interest. 
Because of the bulk relativistic motion of the electrons,
photons impinging on the jet from the side or front are
relativistically boosted in the frame of the electrons, 
enhancing their apparent intensity. This means strong-emission-line sources
are almost certainly dominated by the inverse Compton scattering of external
photons (the external Compton, or EC, mechanism), at least at GeV energies. 
In the weak-lined objects, the BL Lacs, probably the SSC mechanism dominates 
at all gamma-ray energies. 

A few general points about synchrotron and Compton scattering scenarios
should be noted. First, the two SED peaks reflect a characteristic 
electron energy, perhaps where the cooling time, which decreases with
frequency, balances the time-independent escape time. 
Below the synchrotron peak, then, escape could be significant, 
which could explain the very flat infrared spectra sometimes observed,
at wavelengths where the source should
be optically thin. (Flat radio spectra, at longer wavelengths, are
instead caused by optical depth effects.) 

A second interesting point is that, roughly speaking, 
the ratio of peak frequencies in both red and blue blazars,
$\nu_C$/$\nu_s$, appear similar.
(Bear in mind, however, that $\nu_C$ can be
affected by Klein-Nishina effects in the blue blazars, 
and that gamma-ray data are very sparse.)
Assuming this is physically meaningful, the interpretation depends on 
how the gamma-rays are produced at the second peak.
If SSC dominates in all blazars, then $\nu_C/\nu_s \propto \gamma_{peak}^2$
(where $\gamma_{peak}$ is the characteristic energy of electrons radiating
at the synchrotron peak),
and one would conclude that the peak electron energies are 
independent of blazar luminosity. 
If instead the gamma-rays are produced by the EC mechanism in all blazars, 
then $\nu_C/\nu_s \propto \nu_{ext}/B$, where $\nu_{ext}$ is the typical
frequency of the soft seed photon and B is the magnetic field. This would
imply the ratio of $\nu_{ext}/B$, a sort of ratio of accretion energy
to magnetic energy, is roughly constant in all blazars.
More likely, SSC dominates the gamma-ray peak in low-luminosity blazars
and EC the peak in high-luminosity blazars (see \S~\ref{ssec:trends}).
In this case
$\nu_{ext}/B$ in luminous blazars must approximately ``match'' 
$\gamma_{peak}^2$ in less luminous blazars. This might indicate that
electron acceleration, which determines $\gamma_{peak}$ when cooling is 
less important, is related to the magnetic field strength and 
overall source (accretion) luminosity, in a kind of energy balance
constraint.

\subsection{General considerations for emission models}

There remains the question of whether blazar jets are homogeneous or
inhomogeneous. (Of course, jets are unlikely to have uniform electron
densities and magnetic fields on all scales, but at high energies the 
dominant emission region could be approximately homogeneous.)
Perhaps surprisingly, much of the existing multiwavelength data are 
still consistent with a homogeneous model. For example, the lags
between soft and hard X-rays can be explained by energy-dependent
cooling in a homogeneous volume (see \S~\ref{sec:blue}).
However, inhomogeneities may be needed to explain variability amplitudes
even at the highest energies.
For example, at X-ray energies in blue blazars, 
we typically see variations by factors of 2, not factors of 10 or 10\%. 
It is natural to get very large amplitude variations from a single 
acceleration event --- the high-energy emission should be ``on'' or ``off''
because the loss time scales are very short. At the other extreme,
small amplitude variations occur naturally from the superposition 
of a large number of strong flares.
To get the observed, intermediate amplitude variability is 
unnatural in the homogeneous case, and implies instead 
a quiescent flux, at a level comparable to the flare peak intensity. 
This implies that flares are on top of some quiescent
level, and that the flaring region varies with large amplitude, 
which suggests that jets are at least partially inhomogeneous 
(Sikora, Madejski \& Begelman 1997).

If the dominant emission region is homogeneous, 
there must be additional electron acceleration above the 
characteristic $\gamma_{peak}$, for at least two reasons.
First, a sharp cutoff in electron energy should produce
much larger amplitude variations than are observed.
Second, the TeV spectra of Mrk~421 and Mrk~501 extend to 10 TeV, 
perhaps even to 30 TeV in some bright states (Krawczynski, this
volume); if there were
a sharp cutoff at $\gamma_{peak}$, the gamma-ray spectrum should cutoff
sharply as well. 
Thus the acceleration process(es) must produce 
electrons of a characteristic energy $\gamma_{peak}$ {\it and} produce a 
power-law distribution of electron energies above $\gamma_{peak}$.
This might occur because electrons are universally accelerated
to $\gamma_{peak}$ in a blast wave or bulk acceleration process
then further accelerated by shocks, or because they are injected at
high energies and cool to $\gamma_{peak}$ (below $\gamma_{peak}$ cooling 
produces a flat electron energy distribution, flatter still if
escape is important; thus electrons at $\gamma_{peak}$ dominate
the emission).

More fundamental issues are how energy is extracted from the 
central black hole and how it is transported to the radiative zone.  
Much of the extracted energy is eventually manifested as kinetic 
energy in the jet, some fraction of which is then radiated in 
the observed SED (the rest feeds the
extended radio lobes). Very near the black hole, the radiation and
particle densities in blazars would be high enough to efficiently 
generate pairs, even taking relativistic beaming into account. 
The resulting cascades would trap
high-energy radiation in an optically thick zone, contrary to
observation. This means the observed emission must originate in
a radiative zone a fair distance from the black hole, and the
energy transport process is effectively dissipationless.

Sikora (1997) has argued that cold protons ($\gamma\sim1$) are unlikely 
to carry much energy in blazar jets, assuming the energy distributions
of electrons and protons are similar. This is because cold electrons,
which would have an effective Lorentz factor $\sim10$ due to 
the bulk relativistic outflow, should Comptonize ambient UV radiation 
up to $\sim1$~keV. The absence of this so-called ``bulk Compton'' bump 
indicates there are too few cold electrons to transport 
much energy along the jet. Instead, there must be a dissipationless
transport of energy, such as Poynting flux.

\subsection {SED trends with luminosity --- the current blazar paradigm} 
\label{ssec:trends}

Observed blazar SEDs change shape systematically with luminosity, as noted by
Rita Sambruna in her Ph.D. thesis (Sambruna et al. 1996, Sambruna 1997) 
and by Fossati et al. (1997). 
Both synchrotron and gamma-ray SED peaks decrease in frequency
with increasing luminosity. 
At the same time, the ratio between the two SED components, $L_C/L_s$ ---
often called the ``Compton dominance'' --- increases,
from $\sim 1$ or even less for the blue BL Lacs to as much as 
100--1000 for some luminous quasars.
The luminosity in emission lines also increases with bolometric
luminosity; the distinction between BL Lacs and quasars, at an equivalent
width of 5 \AA, is essentially an arbitrary one (Scarpa \& Falomo 1997).

This is not to say that BL Lacs and emission-line blazars are identical, 
or even that their properties overlap strongly, simply that there is no 
clear gap between the populations. A possible exception is the VLBI 
polarization properties, which appear to differ discontinuously between 
BL Lacs and quasars (Gabuzda, this volume). This apparent dichotomy could 
be affected,
however, by differences in intrinsic physical scale sampled at fixed
VLBI resolution. Apart from this uncertain issue, 
the distinction between quasars and BL Lacs occurs at an arbitrary point
(i.e., 5 \AA\ equivalent width),
spanned by a continuous distribution of physical properties. 
Lest this cause confusion, I repeat: 
{\it Of course the average properties of quasars
and BL Lacs are different, and indeed the extremes of each class are 
grossly different! But the point at which one defines a BL Lac
is arbitrary because the properties are continuous.}

The observed trends of blazar properties with luminosity, 
if they reflect an intrinsic physical reality, 
support a self-consistent picture of
equilibrium between electron acceleration and cooling
(Ghisellini et al. 1998; Ghisellini, this volume). 
The basic paradigm is that electrons are accelerated to similar high 
energies in all jets, in an approximate power-law distribution, 
and then they cool on the ambient radiation field. 
Because the local photon densities are higher
in more luminous sources, there is more cooling, and the equilibrium
values of $\gamma_{peak}$ are lower, which is reflected in lower frequencies
of the synchrotron SED peaks (see also Georganopoulos \& Marscher 1998;
Marscher, this volume).

This is an attractive and viable paradigm but is not yet well tested.
The multiwavelength data for existing samples are sparse and not uniform, 
particularly at gamma-ray energies.
To confirm that the two SED components are produced by
the same electrons requires much better multiwavelength monitoring data, 
to find correlated variations and to ``identify'' the electrons
radiating at each wavelength.

Furthermore, the luminosity-dependent paradigm is based 
on blazars for which there are gamma-ray detections, 
and those samples may be biased in the direction of the 
observed SED trends, either through the extra beaming in EC sources
(Dermer 1995) or through variability (Hartman and Wagner, this volume).
That is, the gamma-ray-detected blazars may not be representative of blazars
as a class, nor representative of the average state of the particular blazar.

\section{Multiwavelength variability }
\label{sec:blue}

Because they are very bright at op\-ti\-cal/UV/X--ray wave\-lengths, 
Mrk~421, Mrk~501 and PKS~2155--304 are the best-monitored blazars. 
Each has been the target of multiple well-sampled, long-duration 
monitoring campaigns. For Mrk~421, both the high-energy synchrotron 
emission and the gamma-ray component have been well monitored 
because this blazar is consistently bright at TeV energies. 
Mrk~501 has more recently been well monitored, following a dramatic
TeV and X-ray flare in April 1997.
PKS~2155--304 has been monitored for longer and with better sampling
at optical through X-ray wavelengths
but without simultaneous TeV monitoring (cf. Chadwick et al. 1998).

In all three of these blue blazars, the synchrotron component dominates 
out to at least a few keV. In both Mrk~421 and PKS~2155--304,
the hard X-rays ($\sim 5$~keV) lag soft X-rays ($\sim 0.5$~keV) 
by ~1 hour (Takahashi et al. 1996, Urry et al. 1997;
also Treves, this volume), roughly as expected
for a homogeneous source, where the lag time is roughly the 
radiative time scale at the lower energy.

Of great interest is the cor\-re\-lat\-ed var\-i\-a\-bi\-li\-ty be\-tween syn\-chro\-tron
and Comp\-ton com\-po\-nents at corresponding points on the SED; for
these blue blazars, the spectral peaks are in the X-ray and TeV bands.
Mrk~421 appears to flare roughly
simultaneously in X-rays and TeV but the lag is very poorly constrained
(Macomb et al. 1995, Buckley et al. 1996).
More recent multiwavelength monitoring of Mrk~421 in April 1998, for a full two
weeks with much denser X-ray sampling, should provide better constraints
(Takahashi et al., in prep.).
During a bright flare in 1997 (Catanese et al. 1997, Pian et al. 1998),
Mrk~501 was easily detected at TeV energies and with the RXTE X-ray 
All-Sky Monitor (ASM). The cross-correlation of the two well-sampled
light curves (with at least daily sampling, over several months)
has a highly significant peak at zero lag, 
with an uncertainty of $\sim\pm1$~day (Krawcynski, this volume).
This is the first well-measured X-ray/TeV correlation.

The close correlation of X-rays and TeV emission in blue blazars
is strong evidence for a common production mechanism for both 
synchrotron and gamma-ray SED components. Measuring the lags
accurately between the real SED peaks, a good determination of which 
requires detailed monitoring over a large wavelength range,
should distinguish between Compton scattering models
and alternatives like proton-induced cascades (PIC).

In the PIC model, protons and electrons are accelerated 
to extremely high energies; the primary electrons radiate the low-energy 
SED component via synchrotron;
protons interact with ambient photons to induce pair cascades;
and a subsequent (third?) generation of pairs radiates the high-energy 
SED component, also via synchrotron radiation.
The low-energy radiation should appear first, followed by the
high-energy component, and the relative amplitude of variability
at the two peaks is essentially free.
In contrast, the Compton scattering picture predicts true simultaneity 
and larger amplitude in the gamma-rays, 
at least for flares caused by electron acceleration.

Variability amplitudes are very informative, 
although the observational data at present are somewhat limited. 
This is because, for most GeV blazars, EGRET lacks the dynamic range 
to measure variations by more than a factor of 10 or so, 
and also because the corresponding synchrotron peaks are poorly monitored,
being in the far infrared.
For the TeV blazars, the synchrotron peaks can be well observed but
there are very few sources detected in gamma-rays.
That said, if the multi-epoch SED of 3C279 is typical (Fig.~2),
the gamma-ray component is generally much more variable than the 
synchrotron component. The EGRET data are near the gamma-ray SED peak
in 3C279, which varies in intensity by more than a factor of 30;
above the synchrotron peak, in the optical/UV, the simultaneous
intensity varied by less than a factor of 2 (Wehrle et al. 1998).
Indeed, the gamma-ray variation is more than the square of the 
apparent synchrotron variation (with the caveat that we do not actually
observe the infrared peak). It should go roughly linearly 
in the external Compton model and roughly as the square in the SSC model,
for simple electron acceleration events. 
Clearly additional complexity is present, such as enhanced beaming
in the external Compton case (Dermer 1995) or possibly a change in 
the Doppler factor (Dermer \& Chiang 1998)

The synchrotron peak frequencies in individual objects can 
change dramatically during a flare, increasing by a factor
of $\sim100$ during the strong Mrk~501 flare
(Pian et al. 1998, Catanese et al. 1997; also Pian, this volume).
A corresponding shift in the gamma-ray peak was not seen,
probably because Klein-Nishina effects dominate at the peak.
Such strong spectral hardening in blazars had not previously been 
observed, and it is important to look for these events now
that BeppoSAX provides sufficient hard X-ray sensitivity.
(It remains difficult to observe similar spectral hardening 
in the red blazars because their peaks are often in the far
infrared spectral region, where there are few monitoring data,
perhaps a handful of ISO results slowly coming out.)\\

\noindent
\section{What kind of jets does nature make?}

A prerequisite for understanding jet formation is 
understanding the demographics of jets in radio-loud AGN. 
It is now recognized that blazar jets, as evidenced by their 
range of SEDs, encompass a range of physical states. 
To first order, it appears that
the low luminosity, low kinetic power jets of blue blazars
have systematically higher frequency synchrotron cutoffs and therefore 
higher electron energies, while the high luminosity, high kinetic power 
jets of red blazars have lower electron energies. 
However jets are formed, nature produces a distribution of
intrinsic properties.

Selection effects are strong in blazar samples. 
Not only is their jet emission strongly beamed, 
but the principal selection bands for existing flux-limited samples, 
radio and X-ray, are at well separated points on the SED. Thus it is
no surprise that current blazar samples are highly biased. 
Nonetheless, people seem surprised to learn that the
extrapolation from those samples, to infer
the relative numbers of low-power (blue) and high-power (red) jets,
is uncertain {\it by a factor of $\sim100$}, depending on assumptions.
This is illustrated (for BL Lacs) by the comparison of integral
number counts for radio and X-ray samples (Fig.~3). 
The left panel shows the number counts (Urry, Padovani \& Stickel 1991) 
from five major X-ray samples (open symbols), 
which are primarily blue BL Lacs,
and the ``bivariate'' X-ray counts for the 
1~Jy radio sample (filled triangles)
which are primarily red BL Lacs.
Under the assumption that
the X-ray surveys are an unbiased probe of blazars (i.e., unaffected by 
their red/blue nature), one would conclude that blue BL Lacs are ten
times more numerous than red BL Lacs.

However, under the assumption that the 1~Jy survey is instead unbiased,
one comes to the opposite conclusion 
(Giommi \& Padovani 1994, Padovani \& Giommi 1995).
The right hand panel in Figure~3 shows
the integral radio counts for the 1~Jy sample, combined with the 
``bivariate'' radio counts for the X-ray samples, i.e., the 
X-ray-derived surface density of blue BL Lacs at the observed radio
fluxes for those samples. The solid line shows the extrapolation to lower
radio fluxes of the expected counts for the best-fit unification of
radio galaxies and BL Lac objects (Urry et al. 1991). If the radio survey
were equally sensitive to red and blue BL Lacs, then one would conclude
that the powerful red objects are 10 times more numerous than blue.

Note that the Padovani and Giommi calculation is approximately symmetric 
with respect to assumed SED. They assume a distribution of spectral cutoffs 
toward higher frequency, with the radio survey being unbiased.
One could equally well assume the X-ray surveys are unbiased, and that
there is a distribution of spectral cutoffs toward lower frequency.
The ensuing calculation does indeed reproduce the observed radio counts
(Fossati et al. 1997). {\it Thus the relative numbers of red and blue
blazars remain indeterminate at present.}

Because blazar SED shapes strongly affect surveys at both radio and 
X-ray wavelengths, the truth probably lies somewhere in between 
(see Padovani, this volume). An attempt to estimate
the bolometric luminosity function, thus correcting in a primitive way
for the luminosity effect (but still a ``bivariate'' approach in the
sense that missed objects are still missing), implied a much more similar
space density in the region of overlapping luminosity (Urry \& Padovani
1995). But the fact remains that the question of relative space density
is still largely an 
open one, and requires deeper surveys to solve. The most promising avenue
is to look at deeper flat-spectrum radio samples; the flux limit of the
1~Jy sample is so high that going even a factor of a few deeper should
reveal differences between the extreme X-ray-dominant and radio-dominant
scenarios. 

I have described how, at present, the distribution of jets in nature,
high luminosity red jets versus low power blue jets, 
is not well constrained. Obviously nature makes both types,
and their ratio may change in cosmic time. In particular,
if blazar power/luminosity decreases in cosmic time, as suggested by
global evolutionary properties of AGN, there should be more blue
BL Lacs locally and more red ones at high redshift, more or less consistent
with the different evolutionary properties of the two types.

\begin{minipage}{0.99\textwidth}

\begin{minipage}{0.6\linewidth}
\hspace{-1cm}
\psfig{file=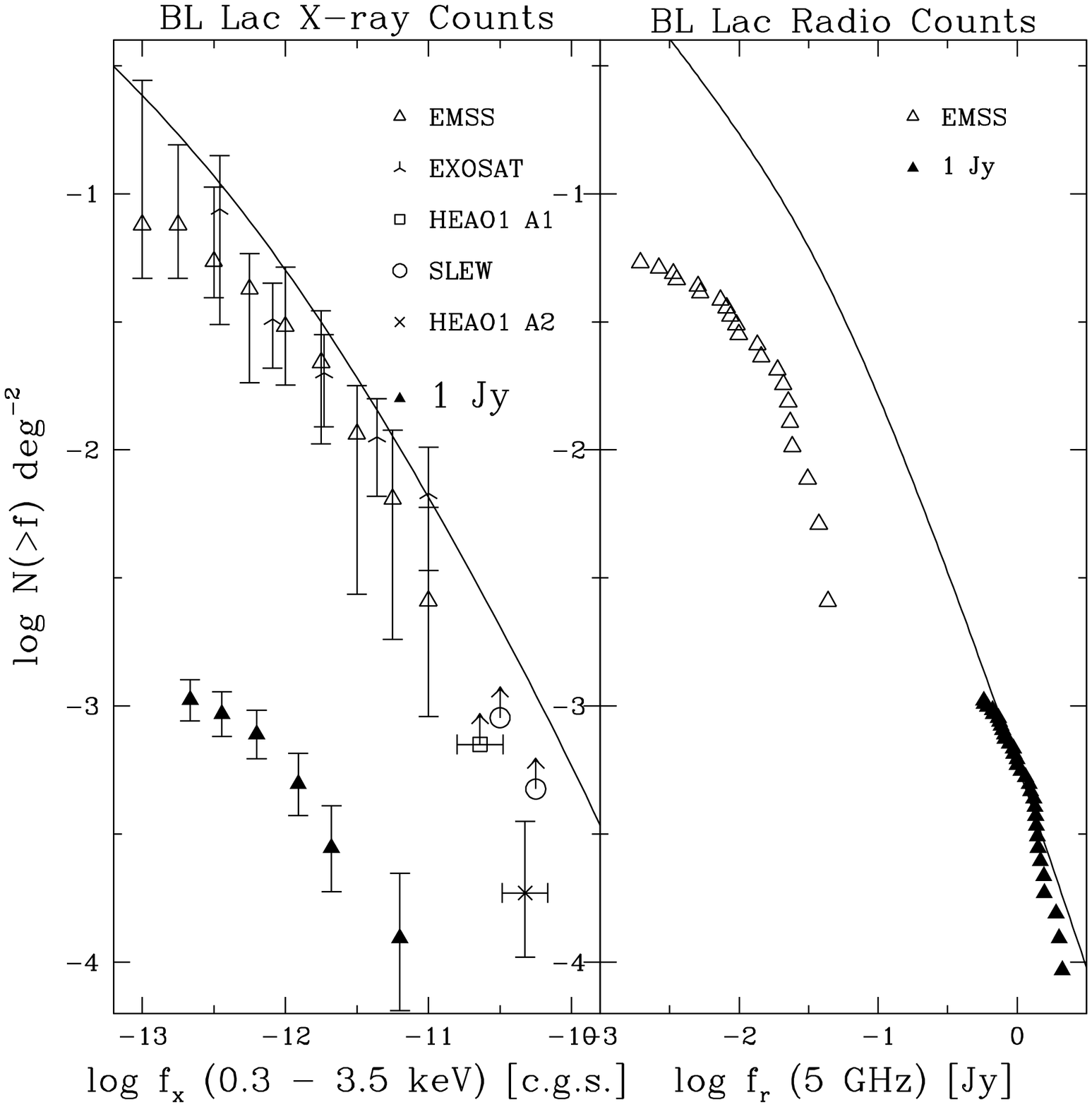,width=1\linewidth}
\end{minipage}  \  \
\hspace{-0.5cm}
\begin{minipage}{0.38\linewidth}
Figure 3. The X-ray and radio number counts for blue (open symbols) and
red (filled symbols) BL Lac objects illustrate the ambiguity over which
population is more numerous. 
{\it Left ---} 
Observed X-ray counts for X-ray-selected samples 
are $\sim10$ times higher 
than the ``bivariate'' X-ray counts for radio-selected samples
(radio-derived surface density combined with observed X-ray fluxes;
Urry et al. 1991).
 
{\it Right ---} 
Beaming model extrapolation (solid line) of 
observed radio counts for radio-selected samples (Urry \& Padovani 1995)
is $\sim10$ times higher than the ``bivariate'' radio counts 
for X-ray-selected samples (X-ray-derived surface density 
combined with the observed radio fluxes).
\label{fig:counts}

\end{minipage}

\begin{minipage}{0.6\linewidth}
\vspace{1cm}
\hspace{-1cm} 
\psfig{file=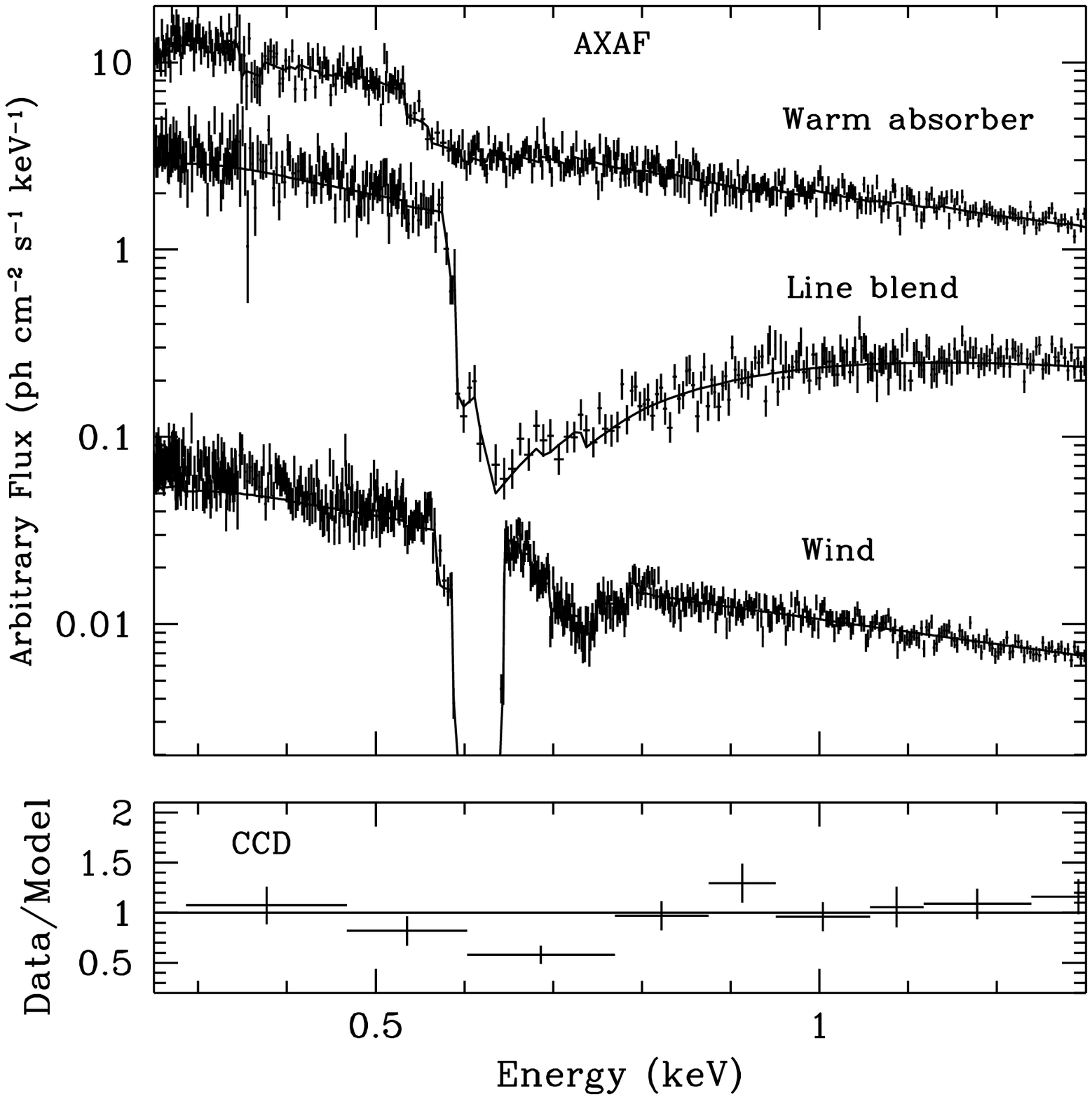,width=1\linewidth}
\end{minipage} \  \
\hspace{-0.5cm} \vspace{1cm}
\begin{minipage}{0.38\linewidth}
Figure 4. Simulated soft X-ray spectra for three viable models of circumnuclear
absorption in blazars, at typical grating resolution. Current low-resolution
CCD data have detected broad absorption clearly but cannot unambiguously 
determine its origin, as shown by the residuals to a simple power law fit 
to ASCA data (bottom panel). AXAF, XMM and/or Astro-E will clearly 
distinguish among the possibilities.
\label{fig:axaf}

\end{minipage}

\end{minipage}

\section{Sub-parsec structure}

Very little is known about the sub-parsec structure of BL Lac objects. 
According to the AGN paradigm, unless structures change drastically 
at low luminosity (and there are few if any hallmarks of such
abrupt change), the centers of BL Lacs should contain the same broad
emission line regions and accretion disks as conventional quasars. 
Clearly at least some BL Lacs have a broad-line region 
(e.g., Corbett et al. 1996, Vermeulen et al. 1995, Stickel et al. 1991), 
as was evident already at the Pittsburgh conference 
(MgII lines seen in BL Lacs; Miller 1978).
These properties have not been surveyed systematically, the blue bump 
and broad lines being hidden in large part by the relativistically boosted 
jet continuum, so the distribution of line or accretion disk luminosities
is only poorly known. It is extremely important to survey the
``thermal'' properties of blazars.

One clue comes from X-ray spectroscopy. At present,
the state of the art --- CCD spectroscopy with resolving power 
of $\sim10$ in the soft X-ray energy range --- suggests there is 
highly ionized circumnuclear material in BL Lac objects, producing
absorption near 0.6~keV (Fig.~4). 
At least three quite different models for the absorber --- 
an outflowing wind, a line blend, and a so-called ``warm absorber'' ---
fit the low-resolution CCD data equally well. With the much higher 
spectral resolution soon available with AXAF, XMM, and Astro-E,
identifying the origin of the X-ray absorbing gas in BL Lacs, 
and perhaps mapping it crudely via variability, 
should be relatively easy. Such data can be used to map the innermost
structure of BL Lac objects.

\section{Host galaxies}

Host galaxies are easily detected in BL Lac objects, as found in 
our large HST survey (Falomo et al. 1997, Urry et al. 1999a,b;
see also Falomo, this volume).
BL Lac host galaxies are large, round, smooth ellipticals,
roughly 1~mag brighter than a typical ($L^*$) elliptical
(Wurtz et al. 1996; Urry et al. 1999b),
fitting well the $\mu_e$-$r_e$ projection of the fundamental plane 
for elliptical galaxies, at the high luminosity end (Urry et al. 1999b).

According to unified schemes, the unbeamed properties of BL Lac objects,
such as host galaxies, should be identical to those of their parent 
population, nominally FRI radio galaxies.
Indeed, BL Lac host galaxy magnitudes are typical of FRIs in the same
redshift range. 
({\it N.B.} Because flux-limited samples of BL Lacs usually span higher 
redshifts than FRIs, as do FRIIs, simple comparisons of host-galaxy 
magnitude distribution, ignoring redshift, can lead to incorrect conclusions.
Another common mistake is to compare, in a KS sense, the observed
distribution of some property of blazars and radio galaxies, such as
host galaxy magnitude. Because of selection effects related to beaming 
and/or the redshift-luminosity correlation in flux-limited samples, 
the shape of the distributions will not necessarily match, even in
intrinsically identical AGN; rather, the parameter values ``match''
if they span the same range.)

Furthermore, both the host galaxy magnitude and the extended radio power
of BL Lacs must match those of the parent population, since the
division between FRI and FRII galaxies appears sharp only in this
two-dimensional parameter space (Owen \& Ledlow 1994).
Figure~5 shows clearly that the extended radio powers and 
host galaxy magnitudes of BL Lac objects overlap those of 
FRI radio galaxies and are unlike FRIIs.
The naive projection of these data in optical magnitude alone,
independent of radio luminosity, has led some to the incorrect conclusion
that FRIIs are a better match for the parent population. This mistake
happens in part because there are no detections of BL Lac host galaxies 
as luminous as the very brightest FRIs, typically the central dominant 
galaxies in rich clusters. Indeed, BL Lacs may avoid the very richest
clusters (Owen, Ledlow \& Keel 1996), but as Figure~5 illustrates, this
must be a second-order issue in the largely correct unification of
FRIs and BL Lacs.

\begin{minipage}{0.5\textheight}
\vspace{1cm}
\centerline{\psfig{file=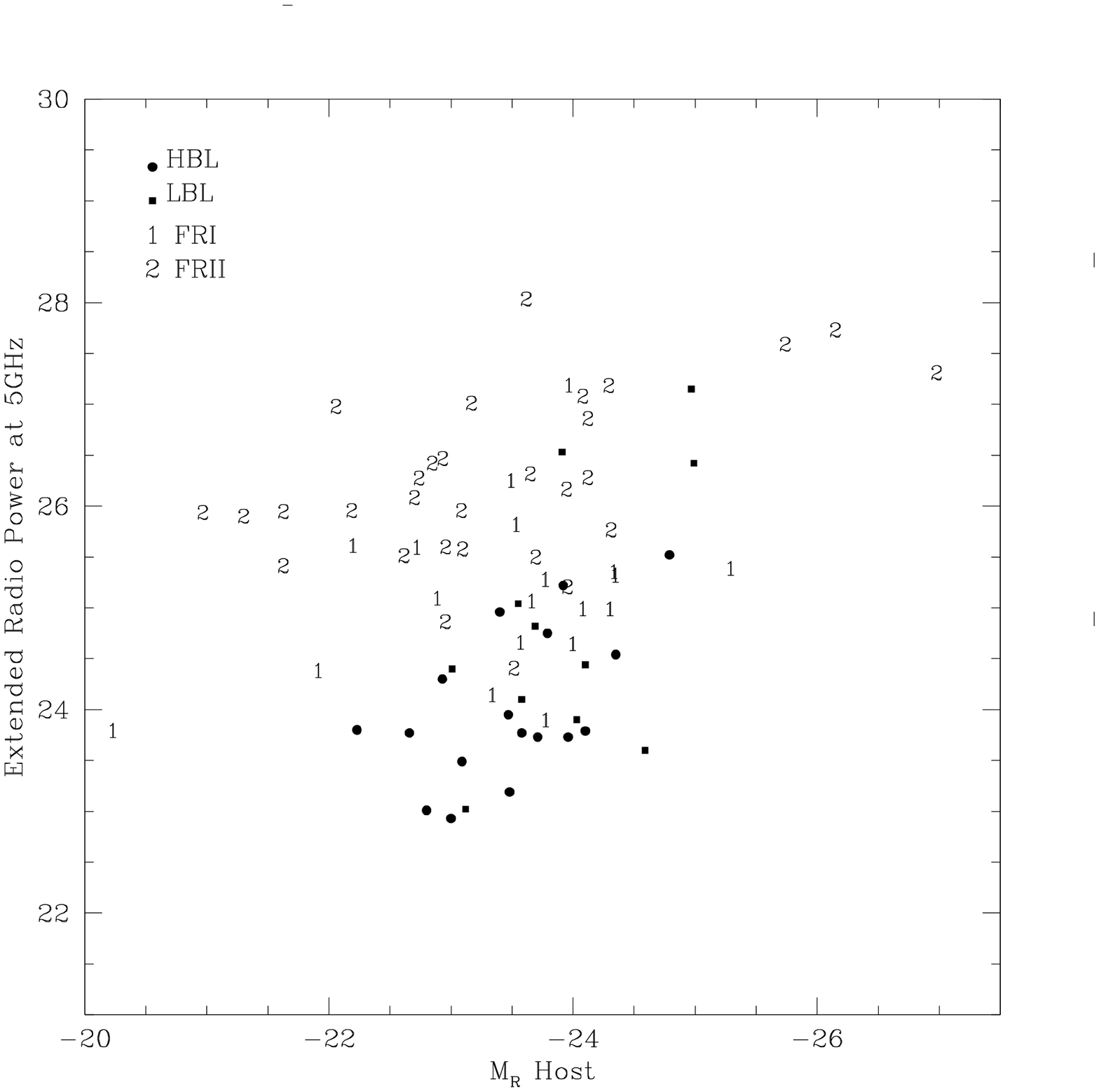,width=0.6\linewidth}}
Figure 5. Comparison of unbeamed properties --- extended radio power
versus host galaxy magnitude --- of BL Lac objects (filled circles and 
squares) and radio galaxies shows that FRIs (labeled ``1'') 
are a much more likely parent population than FRIIs (labeled ``2''). 
The BL Lacs were taken from complete radio- or X-ray flux limited
samples and host galaxy determinations were made using HST data (Urry
et al. 1999b). The FRI/II samples were from the 2~Jy sample (Wall \& Peacock
1985), selected in a similar way to the BL Lac radio samples.
\label{fig:radopt}
\end{minipage}
\normalsize

\section{Radio-loud versus radio-quiet AGN}

Now for some speculation about how studies of
BL Lac objects might inform our larger understanding of AGN. 
A critical issue is the radio-loud/radio-quiet dichotomy,
which essentially translates to why only some AGN appear to
form relativistic jets.
Blazar studies have indicated the importance of selection biases 
in samples flux-limited at wavelengths located in different parts of
the SED. By analogy, is it possible that the distribution of $\alpha_{ro}$
is not intrinsically bimodal, i.e., that there is no sharp physical 
distinction between radio-strong and radio-weak AGN?
The radio flux distribution for PG quasars is not so strongly
bimodal (Kellermann et al. 1989); perhaps beaming effects (as yet
unquantified) could enhance the radio relative to the optical emission
in the so-called radio-loud objects. 
There do exist ``transition'' objects with intermediate values of 
$\alpha_{ro}$, which have been interpreted as radio-quiet quasars 
with significantly beamed cores (Falcke, Patnaik \& Sherwood 1996). 
In deeper radio samples, intermediate values of $\alpha_{ro}$ should
be further filled in; possibly this is seen in the FIRST radio survey 
(Helfand, priv. comm.)

Alternatively, sharp transitions (in\-trin\-sic bi\-mo\-da\-li\-ty) in ra\-dio--loud\-ness 
could result from a critical dependence on jet kinetic power and 
host galaxy mass, analogous to the FRI/II transition (Bicknell 1995).
At very high jet powers, one gets an FRII type source (an 
emission line blazar when aligned); at lower jet power, there is
a sharp transition to diffuse FRI-type jets (a 
BL Lac object when aligned); perhaps at some still lower power, 
the jet cannot form or is quenched so early that one 
gets a radio-quiet object. 

Flux-limited samples can be corrected in a straightforward way for 
objects that have been missed because of their lower fluxes. Thus
we have confidence that they are representative. But such corrections
work only {\it when the excluded objects are intrinsically identical 
to the discovered objects}. If the missing objects instead have a 
different spectral energy distribution,
then their number and properties depend on assumptions about their 
numbers and SEDs --- basically, the correction is a circular (model dependent)
problem. As far as I can see, this problem has not been addressed
explicitly for the radio-loud/radio-quiet issue.

\section{Questions for the future}

The future in blazar research is very rich. Here is a personal list 
of the key questions ahead:

\noindent
{\bf What is the energy production and balance in blazar jets?}
To understand the extraction of energy from the black hole, 
we have first to quantify the energy distribution among particles,
magnetic fields, and photons. A practical near-term issue is identifying
how the gamma-rays are produced. Several years without GeV data loom ahead,
so for the moment progress means detecting more TeV sources; the 
prospects are good, as TeV telescopes rapidly improve and extend
to southern latitudes. 
TeV sources should be monitored also in X-rays, to look for correlated
variability in the two spectral components; ditto for
red objects at optical and GeV energies with GLAST, possibly
sooner with AGILE-type gamma-ray satellites or ground-based
gamma-ray telescopes like MAGIC. 
The goal is to identify the electrons radiating in each component,
and to see if/how they are related. 
PIC models predict more independent variations than EC or SSC models.

\noindent
{\bf What kind of jets does nature make?}
Quantifying the distribution of jet powers --- the intrinsic luminosity 
function of jets, in a global sense --- is an important constraint
on jet formation scenarios. The present uncertainty in the relative 
numbers of blue and red jets can be resolved with deeper samples
(some reported here; e.g., Perlman, Laurent-Muehleisen, this volume).
Given the range of SEDs in blazars, selection effects are strong and
samples flux-limited in any one band offer a necessarily biased and 
incomplete picture. 

\noindent
{\bf How do radio-loud AGN form and evolve?}
The study of host galaxies is a direct probe of the relative evolution 
of nuclei (black holes) and normal galaxies.
The high-redshift blazars turning up in deeper samples (e.g., DXRBS)
are essential for studying the evolution of low-luminosity jets 
(which must be beamed to be seen at high redshift). 
The new and upcoming large telescopes with infrared sensitivity,
including NGST, the VLT, Magellan, Gemini, etc., are ideally suited
to these studies.
Further study of blazar environments is also important.
With its high spatial resolution and excellent sensitivity,
AXAF will be a powerful tool for detecting clusters around BL Lac objects.
This is not only easier than counting galaxies in the optical and constraining
cluster membership with multi-object spectroscopy, but it probes
the diffuse gas, which contains most of the mass, so classification
of richness is more meaningful.

\acknowledgments 
For their hospitality during my sabbatical, I thank the Center for 
Astrophysical Sciences at the Johns Hopkins University and the 
Brera Observatory of Milan. I am grateful to my blazar
colleagues in Italy and elsewhere for stimulating discussions and 
collaborations, and I thank the organizers for hosting an excellent
conference in Turku. This work was supported
in part by NASA grants NAG5-2538 and GO-06363.01-95A, and was based in
part on observations made with the NASA/ESA
Hubble Space Telescope, obtained at the Space Telescope Science
Institute, which is operated by the Association of Universities for
Research in Astronomy, Inc., under NASA contract NAS~5-26555.
{\it P.S. The die-hard blazar man is Dick Miller, the only person to
attend all three BL Lac conferences.}

\end{document}